\def\mnras{MNRAS}

\def\aj{AJ}

\def\apjl{ApJ}
\def\apj{ApJ}
\def\apjs{ApJS}

\def\nat{Nature}
\def\um{$\mu$m}

\documentclass[useAMS,usenatbib]{mn2e}

\usepackage{subfigure}
\usepackage{graphicx}
\begin{document}

\title{An AzTEC 1.1\,mm Survey of the GOODS-N
Field I: Maps, Catalogue, and Source Statistics}

\author[T.~A.~Perera et al.]{T.~A.~Perera,$^1$
E.~L.~Chapin,$^2$
J.E.~Austermann,$^1$ 
K.S.~Scott,$^1$ 
G.W.~Wilson,$^1$ 
\newauthor
M.~Halpern,$^2$
A.~Pope,$^{2,3}$
D.~Scott,$^2$
M.S.~Yun,$^1$
J.~D.~Lowenthal,$^4$
G.~Morrison,$^{5,6}$
\newauthor
I.~Aretxaga,$^7$
J.J.~Bock,$^8$
K.~Coppin,$^9$
M.~Crowe,$^2$
L.~Frey,$^2$
D.H.~Hughes,$^7$
Y.~Kang,$^{10}$ 
\newauthor
S.~Kim,$^{10}$
P.D.~Mauskopf$^{11}$\\
$^1$Department of Astronomy, University of Massachusetts, Amherst, MA 01003, USA\\
$^2$Department of Physics \& Astronomy, University of British Columbia,
6224 Agricultural Road, Vancouver, B\.C\., V6T 1Z1, Canada\\
$^3$Spitzer Fellow; National Optical Astronomy Observatory, 950 North Cherry Avenue, Tucson, AZ 85719, USA\\
$^4$Department of Astronomy, Smith College, Northampton, MA 01063, USA\\
$^5$Institute for Astronomy, University of Hawaii, Honolulu, HI 96822, USA\\
$^6$Canada-France-Hawaii Telescope, Kamuela, HI 96743, USA\\
$^7$Instituto Nacional de Astrof\'{i}sica, \'{O}ptica y
Electr\'{o}nica, Tonantzintla, Puebla, M\'{e}xico\\
$^8$Jet Propulsion Laboratory, California Institute of Technology, Pasadena, CA 91109, USA\\
$^9$Institute for Computational Cosmology, University of Durham, South Road, Durham DH1 3LE, UK\\
$^{10}$Astronomy \& Space Science Department, Sejong University, Seoul,
South Korea\\
$^{11}$Physics and Astronomy, Cardiff University, Wales, UK\\
}

\date{\today}

\pagerange{\pageref{firstpage}--\pageref{lastpage}} \pubyear{2007}

\maketitle

\label{firstpage}

\begin{abstract}
We have conducted a deep and uniform 1.1\,mm survey of the GOODS-N
field with AzTEC on the James Clerk Maxwell Telescope (JCMT).  Here we
present the first results from this survey including maps, the source
catalogue, and 1.1\,mm number-counts.  The results presented here were
obtained from a 245\,arcmin$^2$ region with near uniform coverage to a
depth of 0.96--1.16\,mJy\,beam$^{-1}$.  Our robust catalogue contains
28 source candidates detected with S/N $\ge3.75$, only $\sim$1--2 of
which are expected to be spurious detections.  Of these source
candidates, 8 are also detected by SCUBA at 850\,\um\ in regions where
there is good overlap between the two surveys.  The major
advantage of our survey over that with SCUBA is the uniformity of
coverage.  We calculate number counts using two different techniques:
the first using a frequentist parameter estimation, and the second
using a Bayesian method.  The two sets of results are in good
agreement.  We find that the 1.1\,mm differential number counts are
well described in the 2--6\,mJy range by the functional form $dN/dS =
N' (S'/S)\mathrm{exp}(-S/S')$ with fitted parameters $S' =
1.25\pm0.38$\,mJy and $dN/dS = 300\pm90$\,mJy$^{-1}$\,deg$^{-2}$ at
3~mJy.
\end{abstract}


\section{Introduction}
\label{intro}
 
Identifying and studying the galaxies at high redshift that will
evolve into today's normal and massive galaxies remains a major goal
of observational astrophysics.  Galaxies discovered in deep
sub-millimetre and mm-wavelength surveys
\citep[e.g.][]{Smail1997, Hughes1998, Barger1998, Blain1999A,
Barger1999A, Eales2000, Cowie2002, sescott2002, Webb2003, Borys2003,
Greve2004, Laurent2005} are generally thought to be dominated by
dusty, possibly merger-induced starburst systems and active galactic
nuclei (AGN) at redshifts $z>2$ with star formation rates as high as
SFR $\sim 1000 ~ \mathrm{M}_{\sun}
\mathrm{yr}^{-1}$~\citep{Blain2002}.  The high areal number density of
these sub-mm and mm-detected galaxies (SMGs), combined with their
implied high star formation rates and measured FIR luminosities
\citep[$L_{\mathrm{FIR}}\sim10^{12}
\mathrm{L}_{\sun}$,][]{Kovacs2006,Coppin2008}, makes their estimated contribution
to both the global star formation density and the sub-mm background radiation as
high as 50\% at $z\sim2$ \citep[e.g.,][]{Borys2003,Wall2008}.  Their observed
number counts imply strong evolution between $z=2$ and today
\citep[e.g.][]{sescott2002,Greve2004,Coppin2006}.  The high star formation rates
at early epochs of SMGs generally match the expectation for rapidly forming
elliptical galaxies, a view supported by the high rate of mergers seen locally
in samples of ultra-luminous infrared galaxies \citep[ULIRGs;][]{Borne2000}, which are
plausible local counterparts of distant SMGs.  Together, these characteristics
have led many observers to surmise that SMGs are likely to evolve into the
massive galaxies observed locally \citep[e.g.,][]{Dunlop1994,Smail1997,Bertoldi2007}
 and may hold important clues to the processes
of galaxy and structure formation in general at high redshift.

GOODS-N is one of the most intensively studied extragalactic fields,
with deep multi-wavelength photometric coverage from numerous
ground-based and space-based facilities. These include Chandra in the
X-ray \citep{Alexander2003A}, HST in the optical and NIR
\citep{Giavalisco2004}, Spitzer in the NIR--MIR (Chary et al. in prep.,
Dickinson et al. in prep.), and the Very Large Array in the radio
\citep[][Morrison et al. in prep.]{Richards2000}, as well as highly complete
spectroscopic surveys from ground-based observatories
\citep[e.g.][]{Wirth2004,Cowie2004}.  This field is therefore ideally
suited for deep mm-wavelength studies of SMGs: the extensive coverage
in GOODS-N allows the identification of SMG counterparts in X-ray, UV,
optical, IR, and radio bands, as well as constraints on photometric
redshifts and investigation of SMG power sources and evolution.

\begin{figure}
\centering
\includegraphics[width=\hsize]{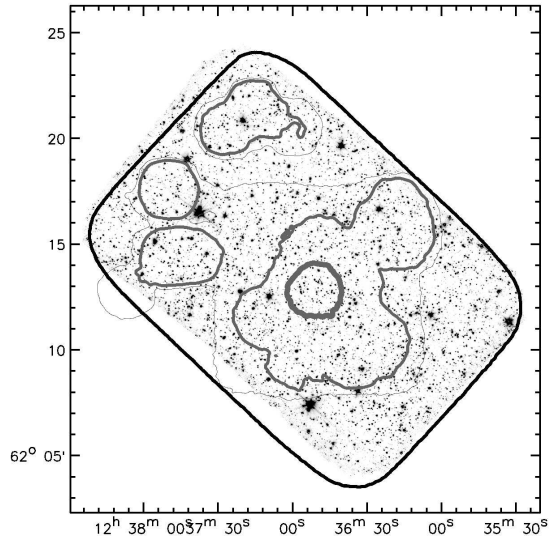}
\caption{AzTEC and SCUBA coverage contours for the GOODS-N region
demonstrates our uniform coverage.  The dark rectangular contour
corresponds to the AzTEC region with a map r.m.s. $\le$1.16\,mJy at
1.1\,mm, the coverage region presented here.  The grey contours,
according to increasing line thickness, are the 850\,\um\ SCUBA
contours for r.m.s. values of 4\,mJy, 2.5\,mJy, and 0.5\,mJy
respectively.  The underlying map is the IRAC 3.6\,\um\ image from the
Spitzer legacy program (Dickinson et al. in prep).  The AzTEC map
represents a significant improvement in the uniformity of coverage at
faint flux levels.}
\label{fig_goodsn_cover}
\end{figure}

Deep mm surveys of blank fields are needed in order to constrain the
faint end of the SMG number counts, while large areal coverage is
required to constrain the bright end.  Together they provide strong
constraints on evolutionary scenarios.  Previous sub-mm surveys of
GOODS-N have been carried out with SCUBA on the JCMT
\citep{Hughes1998,Barger2000,Borys2003,Wang2004,Pope2005}.  The
`Super-map' of the GOODS-N field, which was assembled from all available JCMT
shifts covering the field, contains 40 robust sources at 850\,\um\
down to an average sensitivity of 3.4\,mJy ($1 \sigma$) and covers
200\,arcmin$^2$
\citep{Borys2003,Pope2005}.  However, the r.m.s. is highly non-uniform ranging from
0.4 mJy to 6 mJy (see Fig.~\ref{fig_goodsn_cover}).  That
non-uniformity presents serious complications for comparisons with
multi-wavelength data.

In this paper we present a new 1.1 mm survey of the GOODS-N field made
with AzTEC \citep{Wilson08} at the 15-m James Clerk Maxwell Telescope
(JCMT) on Mauna Kea, Hawaii.  This map is the deepest blank-field
survey undertaken during the AzTEC/JCMT observing campaign, and is one
of the largest, deepest, and most uniform mm-wavelength maps of any
region of the sky.  Our map covers $245$\,arcmin$^2$ and completely
encompasses the $16.5^{\prime}\times10^{\prime}$ {\em Spitzer} GOODS-N
field and all of the previous GOODS-N sub-mm and mm-wavelength fields,
including the original HDF map of
\citet{Hughes1998} and the SCUBA GOODS-N `Super-map'
\citep[indicated in Fig.~\ref{fig_goodsn_cover} here and presented
in][]{Borys2003,Pope2005}.  The large number and high stability of the
AzTEC bolometers has enabled us to produce a map with small variations
in r.m.s., from 0.96--1.16\,mJy, across the 245 min$^2$ field.  This
uniformity is a drastic improvement over the SCUBA GOODS-N `Super-map.'
The sensitivity variations of the AzTEC and SCUBA maps are
compared in Fig.~\ref{fig_goodsn_cover}.

In this work, we extract a catalogue of mm sources from the map and
calculate number counts towards the faint end of the 1.1-mm galaxy
population.  The main results we discuss here were obtained from the
AzTEC data alone; data from other surveys have been used only as tools
to check the quality of our map.  A second paper will address
counterpart identification of our AzTEC sources at other wavelengths
(Chapin et al. in prep.).  We present the JCMT/AzTEC observations of
GOODS-N in
\S~\ref{obs}, data reduction and analysis leading to source
identification in \S~\ref{ana}, properties of our source catalogue in
\S~\ref{sources}, the number counts analysis in
\S~\ref{nc}, the discussion of results in \S~\ref{results}, and the
conclusion in \S~\ref{conclusion}.

\section{AZTEC OBSERVATIONS OF GOODS-N}
\label{obs}
AzTEC is a 144-element focal-plane bolometer array designed for use at
the 50-m Large Millimetre Telescope (LMT) currently nearing completion
on Cerro La Negra, Mexico.  Prior to permanent installation at the
LMT, AzTEC was used on the JCMT between Nov. 2005 and Feb. 2006,
primarily for deep, large-area blank field SMG surveys
\citep[e.g.][Austermann et al. in prep.]{Scott08}.  We imaged the
GOODS-N field at $1.1\,$mm with the AzTEC camera during this
2005--2006 JCMT observing campaign. Details of the AzTEC optical
design, detector array, and instrument performance can be found in
\citet{Wilson08}. Each detector has a roughly Gaussian-shaped beam on
the sky with an 18-arcsec full-width at half-maximum (FWHM).  Given
the beam separation of
22\,arcsec, the hexagonal close-packed array
subtends a ``footprint'' of 5\,arcmin on the sky.  Out of the full
array complement of 144 bolometer-channels, 107 were operational during this
run.

We mapped a 21\,arcmin $\times$ 15\,arcmin area centred on the
GOODS-N field (12$^{\rm h}$37$^{\rm m}$00$^{\rm s}$,
+62$^\circ$13\arcmin00\arcsec) in unchopped raster-scan mode, where
the primary mirror scans the sky at constant velocity, takes a small
orthogonal step, then scans with the same speed in the opposite
direction, repeating until the entire area has been covered. We used a
step size of 9\,arcsec in order to uniformly Nyquist-sample the sky.
We scanned at speeds in the range
60\,arcsec\,s$^{-1}$--180\,arcsec\,s$^{-1}$ as allowed by the fast
time constants of our micro-mesh bolometers, with no adverse
vibrational systematics.  In total, we obtained 50 usable individual
raster-scan observations, each taking 40~minutes (excluding
calibration and pointing overheads). The zenith opacity at 225~GHz is
monitored with the CSO tau meter, and ranged from 0.05--0.27 during
the GOODS-N observations.  This corresponds to 1.1\,mm transmissions
in the range 70--94\%.  A detailed description and justification of
the scan strategy we used can be found in
\citet{Wilson08}.

\section{Data Reduction: from time-streams to source catalogue}
\label{ana}

In this section, we summarise the processing of the AzTEC/GOODS-N
data, which is specifically geared towards finding mm {\em point}
sources.  The data reduction procedure generally follows the method
outlined in \citet{Scott08}, although we emphasise several new pieces
of analysis that were facilitated by the improved depth of this map
over the COSMOS survey.  We begin with the cleaning and calibration of
the time-stream data in \S~\ref{ana_cleaning}, which includes a new
investigation into the sample length over which to clean the data.  In
\S~\ref{ana_map_filter}, we describe the map making process and the
optimal filtering for point sources.  We asses the properties and
quality of the AzTEC/GOODS-N map in \S~\ref{ana_map_qual}.  The depth
of this survey has enabled us to ascertain the degree to which our
data follow Gaussian statistics and detect, directly, a departure from
it at long integration times indicating a component of
signal variance due to source confusion.  The astrometry of the map is
analysed in \S~\ref{ana_stacking}, and we describe the extraction of
sources from the optimally filtered map in
\S~\ref{ana_sources}.

\subsection{Filtering, cleaning, and calibration of time-stream data}
\label{ana_cleaning}

The AzTEC data for each raster-scan observation consists of pointing,
housekeeping (internal thermometry, etc.), and bolometer time-stream
signals.  Because the bolometer data are sampled at 64~Hz, all other
signals are interpolated to that frequency as needed by the analysis.
The raw time-streams of the 107 working bolometers are first despiked
and low-pass filtered at 16\,Hz, as described in \citet{Scott08}.  The
despiked and filtered time-streams are next ``cleaned'' using a
principal component analysis (PCA) approach, which primarily removes
the strong atmospheric signal from the data.  This ``PCA-cleaning''
method was developed by the Bolocam group
\citep{Laurent2005} and later adapted for AzTEC, as described in
\citet{Scott08}.  As explained there, we also generate PCA-cleaned time streams
corresponding to a simulated point source near the field centre, in
order to produce the {\em point-source kernel}, which is used later
for beam-smoothing our maps (see \S~\ref{ana_map_filter}).

In this work we go beyond the analysis in \citet{Scott08} to verify
that we have made good choices with regard to several aspects of the
general cleaning procedure that has been adopted for all of the
existing AzTEC data.  We examine two outstanding questions in
particular: 1) does PCA-cleaning work better than a simple
common-mode subtraction based only on the average signal measured by
all detectors as a function of time? and 2) over what time scale
should each eigenvector projection be calculated in order to give the
best results?

The first question addresses whether simple physical models may be
used in place of PCA-cleaning, where the choice of which modes to
remove from the data is not physically motivated.  We investigate this
by creating a simple sky-signal template as the average of all of the
detectors at each time sample.  We then fit for an amplitude
coefficient of the template to each detector by minimising the r.m.s.
between the scaled template and the actual data. This scaled template
is removed from the bolometer data and we examine the residual signal,
which ideally consists only of astronomical signal and white noise.
We find that this residual signal contains many smaller
detector-detector correlations that are clearly visible in the data
and are dominant compared to the signal produced by astronomical
sources in the map.  The residual time-stream r.m.s.\ from the simple
sky-template subtraction is usually about twice the r.m.s.\ resulting
from PCA cleaning.  This test shows that the simple common-mode
removal technique is insufficient.

In the ``standard'' PCA-cleaning procedure for AzTEC data, outlined in
\citet{Scott08}, the eigenvector
decomposition is performed on each scan ($\sim5$--15~s of data). We
now study which time scales give the best results using a statistical
correlation analysis. We generate a bolometer-bolometer Pearson
correlation matrix using sample lengths that range from a fraction of
a second to tens of minutes (the length of a complete observation).
On the shortest time scales, the correlation coefficients have large
uncertainties due to sample variance (too few samples from which to
make estimates). On time scales corresponding to a single raster-scan
($\sim$5--15~sec), however, the sample variance decreases and a clear
pattern emerges: the strength of the correlations drops off uniformly
with physical separation between the detectors.  The most obvious
trend is the gradient in correlations that we see with detector
elevation, which is presumed to be produced by the underlying gradient
in sky emission.  As the sample length increases, a different pattern
emerges, in which the dominant correlation appears to be related to
the order in which the detectors are sampled by the read-out
electronics, rather than their physical separation.  These
correlations, likely due to electronics-related $1/f$ drifts, are
effectively removed when using scan-sized sample lengths (5--15\,s) as
well, since they appear as DC baseline differences on these short
time-scales.  These results verify that scan-sized sample lengths
produce the best results as they provide a sufficient number of
samples on short enough time scales.

After PCA-cleaning the bolometer signals, we apply a calibration
factor to convert the bolometers' voltage time-streams into units of
Jy per beam.  Details of this procedure are given in
\citet{Wilson08}. The total error on the calibrated signals (including
the error on the absolute flux of Uranus) is 11\%.

\subsection{Map-making and optimal filtering}
\label{ana_map_filter}

The map-making process used to generate the final optimally filtered
AzTEC/GOODS-N map is identical to that used in \citet{Scott08}, and
the reader is directed to that paper for the details of this process,
which we briefly summarise below.

We first generate maps for each of the 50 individual raster-scan
observations separately by binning the time-stream data onto a
3\arcsec~$\times$~3\arcsec\ grid in RA-Dec which is tangent to the
celestial sphere at (12$^\mathrm{h}$37$^\mathrm{m}$00$^\mathrm{s}$,
+62$^\circ$13\arcmin00\arcsec).  We chose the same tangent point and
pixel size as that used for the SCUBA map of GOODS-N
\citep[see for example][]{Pope2006} so that the two maps can easily be
compared in a future paper.  We find that this pixel size provides a
good compromise between reducing computation time, while sampling with
high resolution the 18-arcsec FWHM beams.  Individual signal maps and
their corresponding weight maps for each observation are created as
described in
\citet{Scott08}, along with kernel maps that reflect how a faint point
source is affected by PCA-cleaning and other steps in the
analysis. Next, we form a single ``co-added'' signal map from the
weighted average over all 50 individual observations.  An averaged
kernel map is also created in a similar way.  The total weight map is
calculated by summing the weights from individual observations, pixel
by pixel.  As described by \citet{Scott08} we also generate 100 noise
realization maps corresponding to the co-added map.

We then use a spatial filter to beam-smooth our map using the
point-source kernel, by optimally weighting each spatial-frequency
component of this convolution according to the spatial power spectral
density (PSD) of noise-realization maps.  Details of this optimal filter can
also be found in \citet{Scott08}.

\subsection{Map quality: depth, uniformity, point-source response, and noise integration}
\label{ana_map_qual}

\begin{figure*}
\centering
\includegraphics[width=\hsize]{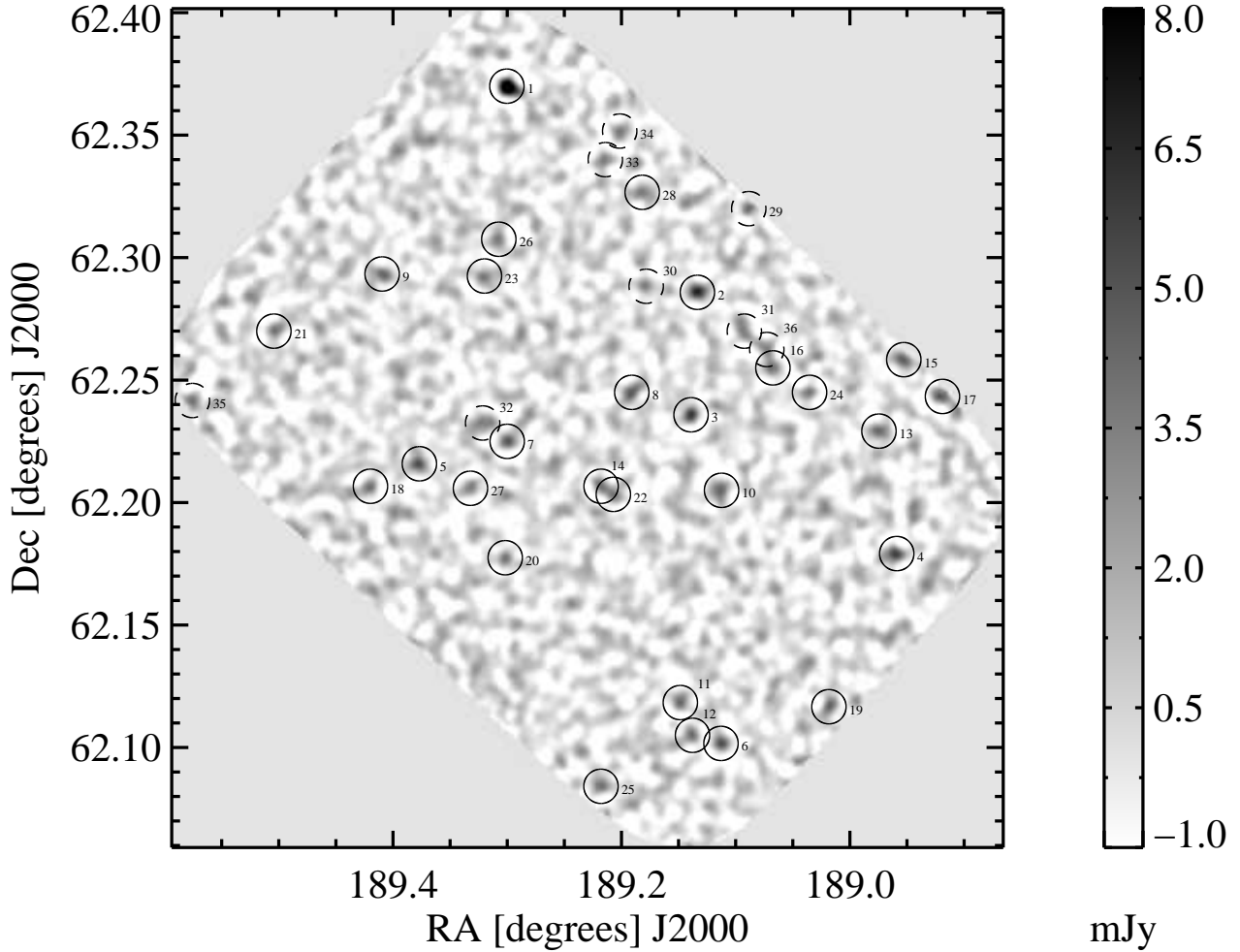}
\caption{AzTEC/GOODS-N signal map with the 36 S/N$\geq$3.5 source
candidates circled.  Information about these source candidates is
given in Table~\ref{table_sources}.  Here and in that table, source
candidates are numbered in decreasing order of S/N.  The source
candidates marked with dashed-line circles do not belong to the
robust sub-list, indicated by a horizontal line in
Table~\ref{table_sources}.  The map has been trimmed to show only the
70\% coverage region (245\,arcmin$^2$).}
\label{fig_source_map}
\end{figure*}
The final co-added, optimally filtered signal map for the GOODS-N
field is shown in Fig.~\ref{fig_source_map}.  Of the
315-arcmin$^2$ solid angle scanned by the telescope boresight
during our survey, we expect $\sim$250\,arcmin$^2$ to be imaged
uniformly by the complete AzTEC array.  We identify this region by
imposing a coverage cut.  We find that weights within 70\% of the
central value occur in a contiguous region of 245\,arcmin$^2$.  The
map of Fig.~\ref{fig_source_map} has been trimmed to only show this
region.  Much of the analysis presented here is limited to this
region, which we will henceforth refer to as the the ``70\% coverage
region.''  The 1-$\sigma$ flux-density error estimates in the trimmed
map range from 0.96~mJy\,beam$^{-1}$ in the centre to
1.16~mJy\,beam$^{-1}$ at the edges.

\begin{figure}
\centering
\includegraphics[width=2.3in,angle=90]{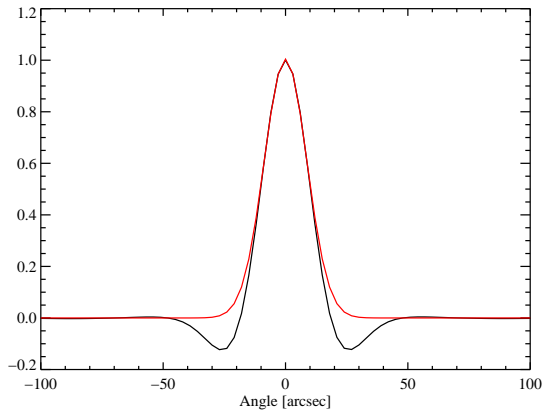}
\caption{Cross section of the point-source kernel.  The Gaussian that best fits
 the inner $R = 10$\,arcsec region is shown in the lighter shade and
 has a FWHM of 19.5\,arcsec.  The negative ring around the centre and
 other peripheral features (not visible here) are induced by
 PCA-cleaning as well as the optimal filter.}
\label{fig_kernel}
\end{figure}
We also run the co-added kernel map through the same filtering process
as the signal map.  The resulting filtered kernel map, whose profile
is shown in Fig.~\ref{fig_kernel}, is our best approximation of the
shape of a point source in the co-added, filtered signal map.  As
demonstrated in
\S~\ref{ana_stacking}, our pointing jitter/uncertainty has a
sub-2-arcsec characteristic scale; this will have little impact on the
kernel shape and therefore is not included in generating the kernel
map. The negative troughs around the central peak are due to a
combination of array common-mode removal in the PCA-cleaning and
de-weighting of longer spatial wavelength modes by the optimal filter.
The point-source kernel also has radial scan-oriented features, or
``spokes,'' due to PCA cleaning that are $<$0.1\% of the kernel
amplitude.  The directions of these spokes would vary across the map
as the scan angle changes with RA-Dec.  Therefore, the kernel map
accurately reflects these spokes only for point sources near the
centre of the field.  However, because it is difficult to analytically
model a point source (through PCA cleaning and optimal filtering) and
because the radial features are very faint, we use the kernel map as a
point source-template for injection of sources in the simulations
described later.

\begin{figure}
\centering
\subfigure{
\hspace{.17in}
\includegraphics[width=2.4in,angle=90]{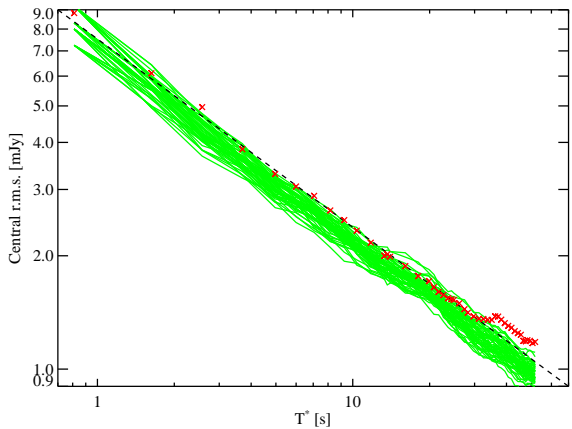}}
\caption{Behaviour of the signal map's r.m.s. (crosses), as well as the
r.m.s. of 100 separate noise realisations (collection of curves), as a
function of the mean effective integration time $T^{*}$ spent on each
3-arcsec central pixel of map.  The dashed curve shows the 1/$\sqrt{T^*}$
relationship expected in the absence of systematics and astronomical
signal. This demonstrates how the map noise averages down with the
successive addition of more observations. The ``flattening'' of the
central r.m.s. at
large $T^*$ in the signal map, compared to the noise maps, is due to
astronomical signal.  The fluctuations of this curve at large $T^*$ are
simply due to noise in the r.m.s. itself, as re-ordering observations
gives similar features near the same region.}
\label{fig_quietness_plot}
\end{figure}

Because this GOODS-N survey is the deepest blank-field survey
conducted thus far with AzTEC on the JCMT, we demonstrate in
Fig.~\ref{fig_quietness_plot} how the map noise averages down with the
successive co-addition of individual observations.  The central
200\arcsec$\times$200\arcsec\ region of the signal map and the noise
realisation maps are used for this calculation.  The x-axis represents
the average weight of a 3-arcsec pixel in this region prior to
filtering.  A scale factor converts this raw weight to an effective
time, $T^{*}$, so that the final effective time equals the final
integration time devoted to an {\em average} 3-arcsec pixel in this
central patch.  Thus, the increment in $T^{*}$ gained with the
addition of an individual observation is the effective integration
time contributed by that particular observation to the central region.
The $i$th y-axis value is calculated by co-adding (averaging)
individual signal maps from observations 1 through $i$, then applying
the optimal filter, and finally taking the standard deviation of this
co-added, filtered map in the central region.  The crosses represent
the signal map.  The 100 curves shown in a lighter shade are
calculated by carrying out the same process on 100 noise realisations.
In the absence of systematics or astronomical signal, we expect all
curves to scale as $1/\sqrt{T^{*}}$, in accordance with Gaussian
statistics, as indicated by the dashed line.  At higher $T^{*}$, we
may expect a slight steepening in all curves because later
co-additions better reflect our assumptions of circular symmetry (in
the optimal filtering process) as we add more scan directions to the
mix.  However, this effect appears to be unmeasurably small in our
data.

While the noise realisations follow the $1/\sqrt{T^{*}}$ trend, the
signal map initially follows it but flattens near the point where
${\sim}20$--30~s of effective time is spent on a 3-arcsec pixel.
Switching the order in which signal maps are co-added does not alter
this trend or the noisy behaviour of these points at large $T^{*}$.
Therefore, we conclude that: 1) single individual observations yield
maps that are consistent with our noise realisations; 2) map features
that do not survive scan-by-scan ``jack-knifing,'' presumably
astronomical signal due to source confusion, prevent the signal map's
r.m.s. from improving as $1/\sqrt{T^{*}}$; and 3) the fact that noise
realisations continue to follow this trend indicates that we are far
from a systematics floor due to atmospheric or instrumental effects,
even at the highest $T^{*}$.

\subsection{Astrometry calibration}
\label{ana_stacking}

The pipeline used to produce this map of GOODS-N interpolates pointing
offsets inferred from regular observations of pointing calibrators
interspersed with science targets \citep{Wilson08,Scott08}. In order
to verify the quality of this pointing model for GOODS-N, both in an
absolute sense, and in terms of small variations between passes, we
compare the AzTEC map with the extremely deep 1.4\,GHz VLA data in
this field
\citep[][Morrison et al. in prep.]{Richards2000}.
The radio data reduction and source list used here is the same as that of
\citet{Pope2006}, with a 1-$\sigma$ noise of $\sim$5.3\,$\mu$Jy at the
phase centre. The catalogue is constructed with a 4-$\sigma$ cut, and
has positional uncertainties $\sim0.2^{\prime\prime}$
(Morrison et al. in prep.).

We stack the signal in the AzTEC map at the positions of radio sources
to check for gross astrometric shifts in the AzTEC pointing model, as
well as any broadening in the stacked signal which may indicate
significant random offsets in the pointing between visits.  A more
detailed comparison between the mm and 1.4\,GHz map is presented in
(Chapin et al. in prep.) to assist with the MIR/NIR identifications of
individual AzTEC SMGs, and the production of radio--NIR SEDs.

The stack was made from the 453 1.4\,GHz source positions that are
within the uniform noise region of the AzTEC map.  As in
\citet{Scott08} we check for an astrometric shift and broadening by
fitting a simple model to the stacked image, which consists of an
astrometric shift ($\delta$RA, $\delta$Dec) to the ideal point source
kernel, convolved with a symmetric Gaussian with standard deviation
$\sigma_{\rm p}$. This Gaussian represents our model for the random
pointing error in the AzTEC map.  We determine maximum likelihood
estimates $\delta$RA $= 0.2^{\prime\prime}$, $\delta$Dec $=
-0.9^{\prime\prime}$, and $\sigma_{\rm p}=0.6^{\prime\prime}$.  The
expected positional uncertainty (in each coordinate) for a point
source with a purely Gaussian beam is approximately
$0.6\times$FWHM/(S/N) \citep[see the Appendix in][]{Ivison2007} where
the FWHM is 18\,arcsec in our case.  The S/N of our stack is
approximately 10, so the expected positional uncertainty is
${\sim}\,1^{\prime\prime}$.  Therefore the total astrometric shift
measured by the fitting process, $0.9^{\prime\prime}$, is consistent
with the hypothesis that there is {\em no} significant underlying
shift. We also note that the $\chi^2$ function for this fit is
extremely shallow along the $\sigma_{\rm p}$ axis, so although the
minimum occurs at $0.6^{\prime\prime}$, it is not significantly more
likely than $0^{\prime\prime}$.  We therefore conclude from this
analysis that there is no significant offset, nor beam broadening
caused by errors in the pointing model.

\subsection{Source finding}
\label{ana_sources}

\begin{figure}
\centering
\includegraphics[width=2.4in,angle=90]{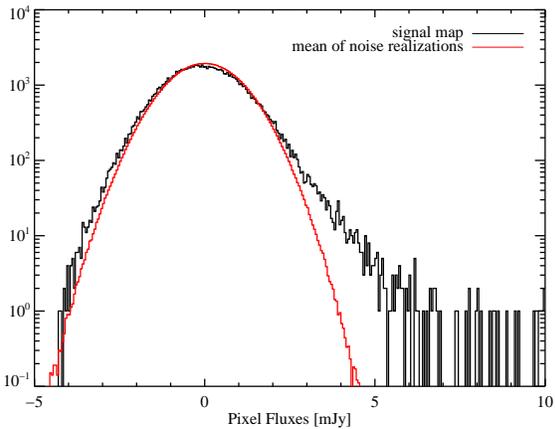}
\caption{Pixel flux histogram of the final signal map in a dark shade and
the average pixel flux histogram made from 100 noise realizations in a
lighter shade.  The positive tail and smaller negative excess in the
signal map is due to the presence of point sources.}
\label{fig_pix_histograms}
\end{figure}
To investigate the presence of astronomical sources in our map, we
plot in Fig.~\ref{fig_pix_histograms} a histogram of pixel fluxes in
the 70\% coverage region of the field.  Also shown in a lighter shade
is the average pixel histogram made from the 100 noise-realization
maps.  The noise histogram can be modelled well by a Gaussian centred
on $0\,$mJy with a standard deviation of 1.0\,mJy.  The obvious excess
of large positive pixel values and the small excess of negative values
in the signal map are caused by the presence of sources.

To identify individual point sources, we first form a S/N map by
multiplying the final (i.e.\ co-added and filtered) signal map by the
square-root of the weight map.  We then identify local maxima in this
S/N map with S/N $\geq 3.5$.  There are 36 local maxima that meet this
condition in the 70\% coverage region of the field.  Our analysis of
these source candidates is simplified because no pair of them are
close enough to significantly alter each other's recovered flux
densities ($>$36\,arcsec apart in each case).  We have evidence that
AzGN01 is a blend of two sources.  However, since this knowledge is
not based on AzTEC data alone, we defer a detailed discussion of that
source for the second paper of this series (Chapin et al. in prep.).

The final signal map and these source candidates
are shown in Fig.~\ref{fig_source_map}.  Table~\ref{table_sources}
lists details of all the AzTEC/GOODS-N $\geq3.5$-$\sigma$ source
candidates, including their locations, measured fluxes, S/N, and
additional quantities which are defined below.  The source positions
are given to sub-pixel resolution by calculating a centroid for each
local maximum based on nearby pixel fluxes.  Sources with clear
counterparts in the SCUBA map of GOODS-N
\citep{Borys2003,Pope2005} are highlighted in
Table~\ref{table_sources}.

\begin{table*}
\centering
\begin{tabular}{l|c|c|c|c|c|c}
\hline
Source ID & RA (J2000) & Dec (J2000) & 1.1\,mm flux [mJy] & source S/N & de-boosted
flux [mJy] & non-positive PFD integral \\
\hline
\hline
AzGN01$^S$ & 12:37:12.04 & 62:22:11.5 & 11.45$\pm$0.99 & 11.58 & 10.69$^{+0.94}_{-1.12}$ & 0.000 \\
AzGN02 & 12:36:32.98 & 62:17:09.4 &  6.84$\pm$0.97 &  7.03 &  5.91$^{+1.02}_{-1.00}$ & 0.000 \\
AzGN03$^S$ & 12:36:33.34 & 62:14:08.9 &  6.23$\pm$0.97 &  6.43 &  5.35$^{+0.94}_{-1.08}$ & 0.000 \\
AzGN04 & 12:35:50.23 & 62:10:44.4 &  5.76$\pm$1.01 &  5.71 &  4.69$^{+1.06}_{-1.06}$ & 0.000 \\
AzGN05 & 12:37:30.53 & 62:12:56.7 &  5.21$\pm$0.97 &  5.38 &  4.13$^{+1.08}_{-0.98}$ & 0.000 \\
AzGN06 & 12:36:27.05 & 62:06:06.0 &  5.28$\pm$1.00 &  5.29 &  4.13$^{+1.12}_{-1.00}$ & 0.000 \\
AzGN07$^S$ & 12:37:11.94 & 62:13:30.1 &  5.04$\pm$0.97 &  5.21 &  3.95$^{+1.08}_{-0.98}$ & 0.000 \\
AzGN08$^S$ & 12:36:45.85 & 62:14:41.9 &  4.94$\pm$0.97 &  5.09 &  3.83$^{+1.08}_{-1.00}$ & 0.000 \\
AzGN09$^S$ & 12:37:38.23 & 62:17:35.6 &  4.50$\pm$0.97 &  4.63 &  3.39$^{+1.02}_{-1.10}$ & 0.003 \\
AzGN10 & 12:36:27.03 & 62:12:18.0 &  4.46$\pm$0.97 &  4.60 &  3.35$^{+1.02}_{-1.10}$ & 0.003 \\
AzGN11 & 12:36:35.62 & 62:07:06.2 &  4.44$\pm$0.98 &  4.53 &  3.27$^{+1.08}_{-1.08}$ & 0.004 \\
AzGN12 & 12:36:33.17 & 62:06:18.1 &  4.32$\pm$0.99 &  4.39 &  3.07$^{+1.12}_{-1.08}$ & 0.008 \\
AzGN13 & 12:35:53.86 & 62:13:45.1 &  4.30$\pm$0.99 &  4.36 &  3.07$^{+1.10}_{-1.12}$ & 0.008 \\
AzGN14$^S$ & 12:36:52.25 & 62:12:24.1 &  4.18$\pm$0.97 &  4.31 &  2.95$^{+1.10}_{-1.08}$ & 0.009 \\
AzGN15 & 12:35:48.64 & 62:15:29.9 &  4.76$\pm$1.12 &  4.26 &  3.23$^{+1.26}_{-1.32}$ & 0.016 \\
AzGN16$^S$ & 12:36:16.18 & 62:15:18.1 &  4.12$\pm$0.97 &  4.23 &  2.89$^{+1.08}_{-1.14}$ & 0.013 \\
AzGN17 & 12:35:40.59 & 62:14:36.1 &  4.75$\pm$1.13 &  4.20 &  3.23$^{+1.24}_{-1.42}$ & 0.020 \\
AzGN18 & 12:37:40.80 & 62:12:23.3 &  4.09$\pm$0.97 &  4.20 &  2.79$^{+1.16}_{-1.08}$ & 0.014 \\
AzGN19 & 12:36:04.33 & 62:07:00.2 &  4.54$\pm$1.09 &  4.15 &  3.07$^{+1.20}_{-1.36}$ & 0.022 \\
AzGN20$^N$ & 12:37:12.36 & 62:10:38.2 &  4.01$\pm$0.97 &  4.14 &  2.79$^{+1.08}_{-1.16}$ & 0.016 \\
AzGN21 & 12:38:01.96 & 62:16:12.6 &  3.99$\pm$0.99 &  4.05 &  2.65$^{+1.16}_{-1.16}$ & 0.023 \\
AzGN22$^N$ & 12:36:49.70 & 62:12:12.0 &  3.81$\pm$0.97 &  3.93 &  2.55$^{+1.08}_{-1.24}$ & 0.030 \\
AzGN23 & 12:37:16.81 & 62:17:32.2 &  3.75$\pm$0.97 &  3.88 &  2.39$^{+1.16}_{-1.18}$ & 0.035 \\
AzGN24$^S$ & 12:36:08.46 & 62:14:41.7 &  3.77$\pm$0.98 &  3.86 &  2.39$^{+1.18}_{-1.20}$ & 0.038 \\
AzGN25 & 12:36:52.30 & 62:05:03.4 &  4.19$\pm$1.09 &  3.85 &  2.55$^{+1.32}_{-1.42}$ & 0.050 \\
AzGN26 & 12:37:13.86 & 62:18:26.8 &  3.70$\pm$0.97 &  3.82 &  2.39$^{+1.10}_{-1.28}$ & 0.041 \\
AzGN27$^N$ & 12:37:19.72 & 62:12:21.5 &  3.68$\pm$0.97 &  3.81 &  2.31$^{+1.16}_{-1.22}$ & 0.043 \\
AzGN28 & 12:36:43.60 & 62:19:35.9 &  3.68$\pm$0.98 &  3.76 &  2.31$^{+1.14}_{-1.30}$ & 0.050 \\
\hline
AzGN29 & 12:36:21.14 & 62:19:12.1 &  4.17$\pm$1.13 &  3.70 &  2.39$^{+1.34}_{-1.64}$ & 0.077 \\
AzGN30 & 12:36:42.83 & 62:17:18.3 &  3.58$\pm$0.97 &  3.69 &  2.13$^{+1.20}_{-1.26}$ & 0.059 \\
AzGN31 & 12:36:22.16 & 62:16:11.0 &  3.58$\pm$0.97 &  3.68 &  2.13$^{+1.20}_{-1.28}$ & 0.061 \\
AzGN32 & 12:37:17.14 & 62:13:56.0 &  3.56$\pm$0.97 &  3.67 &  2.13$^{+1.18}_{-1.28}$ & 0.061 \\
AzGN33 & 12:36:51.42 & 62:20:23.7 &  3.54$\pm$0.98 &  3.63 &  2.13$^{+1.12}_{-1.40}$ & 0.069 \\
AzGN34 & 12:36:48.30 & 62:21:05.5 &  3.65$\pm$1.02 &  3.59 &  2.13$^{+1.16}_{-1.50}$ & 0.080 \\
AzGN35 & 12:38:18.20 & 62:14:29.8 &  4.02$\pm$1.12 &  3.59 &  2.13$^{+1.32}_{-1.68}$ & 0.096 \\
AzGN36 & 12:36:17.38 & 62:15:45.5 &  3.41$\pm$0.97 &  3.50 &  1.87$^{+1.16}_{-1.40}$ & 0.091 \\
\hline
\end{tabular}
\caption{Source candidates in AzTEC/GOODS-N with S/N$\geq$3.5 ordered
according to S/N.  The horizontal line between AzGN28 and AzGN29
represents our threshold for source robustness, as explained in
\S~\ref{sources_fdr}.  The last two columns are defined in
\S~\ref{sources_deboost}.  The superscripts $S$ and $N$ highlight sources
in our robust sub-list that lie within the considered SCUBA region
(where the 850-\um\ r.m.s.\ is $<$2.5\,mJy).  The sources denoted by
$S$ have robust detections at 850\,\um\ within 12\,arcsec of the given
positions while the sources denoted by $N$ do not (Chapin et al. in
prep.).}
\label{table_sources}
\end{table*}

\section{The AzTEC/GOODS-N source catalogue}
\label{sources}

As evident from Table~\ref{table_sources}, the number of source
candidates increases rapidly with decreasing S/N.  However, if we use
a S/N threshold to make a sub-list of the sources in
Table~\ref{table_sources}, the false positives contained in such a
list will also increase with lower S/N thresholds.  Our aim here is to
find a S/N threshold above which $\ga$95\% of source candidates are,
on average, expected to be true sources.  This is a practical choice
aimed at maximising the number of sources recommended for follow-up
studies (the subject of Chapin et al. in prep.) in a way that limits
the effect of false detections on any conclusions drawn.  The
horizontal line in Table~\ref{table_sources} below source AzGN28
(S/N$\geq$3.75) marks the cut-off of the sub-list that we expect will
satisfy our robustness condition.  We first explain in
\S~\ref{sources_fdr} the analysis of false detection rates (FDRs) that
yields this threshold.  In that section, we go beyond previous FDR
treatments for AzTEC \citep{Scott08} and derive some general results
about FDRs that are applicable to (sub)mm surveys in general.

Next, we explain in \S~\ref{sources_deboost} the last two columns of
Table~\ref{table_sources} which contain a re-evaluation of source flux
densities and an assessment of the relative robustness of our source
candidates.  Then, in \S~\ref{sources_completeness}, we discuss the
survey completeness and present a brief consistency check of our
source candidates against SCUBA detections at 850\,\um.

\subsection{False detection rates}
\label{sources_fdr}

Two obvious methods for estimating the false detection rate (FDR) of a
survey are to run the source finding algorithm on: 1) simulated noise
realization maps; or 2) the {\em negative} of the observed signal map.
For several S/N thresholds, Table~\ref{table_fdr} lists the number of
source candidates in the actual map (row 1), the average number of
``sources'' found in simulated pure noise realizations (row 2), and
the number of ``sources'' in the negative of the actual map (row 4).
When using the map negative, regions within 36\,arcsec of a bright
positive source were excluded in order to avoid their ``negative
ring'' (see Fig.~\ref{fig_kernel}).
\begin{table}
\centering
\begin{tabular}{l|c|c|c|c}
\hline
Source Threshold & 3.5-$\sigma$ & 3.75-$\sigma$ & 4-$\sigma$ &
5-$\sigma$ \\
\hline
\hline
Sources Detected & 36 & 28 & 21 & 8 \\
Pure-noise FDR & 4.32 & 1.69 & 0.68 & 0.01 \\
Best-fit-model FDR & 2.65 & 1.13 & 0.42 & 0.00\\
\hline
Negative FDR & 6 & 4 & 4 & 0 \\
Pure-noise Negative FDR & 4.55 & 1.58 & 0.33 & 0.00\\
Best-fit-model Negative FDR & 5.96 & 2.85 & 1.16 & 0.04\\
\hline
\end{tabular}
\caption{The number of source candidates passing a given S/N threshold
in the actual map are indicated in row 1.  Several methods for
determining the false detection rates (FDRs) were explored.
``Pure-noise'' refers to averages computed over 100 noise-realization
maps.  ``Best-fit-model'' corresponds to averages from 100
noise+source realization maps using the best fit model of
\S~\ref{nc_parametric}.  We have settled on the values of row 2 as our
nominal FDRs because they give a conservative overestimate, as
explained in the text.}
\label{table_fdr}
\end{table}

We conclude that these two estimates of the FDR are not very accurate
for our maps.  Because of the high number density of SMGs in the sky
compared to our beam size, every point of the map is in general
affected by the presence of sources.  This source confusion causes the
simple FDR estimates above to be inaccurate.  In particular, there are
equal numbers of negative and positive ``detections'' in noise
realizations to within the statistical error of our noise simulations,
as indicated by rows 2 and 5.  However, the presence of sources skews
this balance in the actual map, making the false negatives rate higher
than the pure-noise numbers and the false positives rate (what we are
after) lower than the pure-noise numbers.

Both these effects can be understood by considering the following
hypothetical construction: a noise-less AzTEC map of the sky
containing many point sources, all with the shape of the point-source
kernel.  Because each kernel has a mean of zero, such a map would have
an excess of negative valued pixels over positive valued pixels (about
70\% to 30\%) to counter the high positive values near the centre of
the kernel (see \ref{fig_kernel}).  When noise that is symmetric
around zero is ``added'' to such a map, this small negative bias will
cause a larger number of high-significance negative excursions in that
sky map compared to a map containing just the symmetric noise.  The
pixel flux histogram of the actual map, shown in
Fig.~\ref{fig_pix_histograms} (darker shade), also shows evidence of
this effect through its negatively shifted peak as well as the excess
of negative pixels in comparison with pure noise realizations
(lighter-shade histogram).  This small negative bias, in pixels that do
not lie atop a source peak, also explains why there are fewer
high-significance false positives in an actual sky map compared to a
pure noise map.

To verify our reasoning, we generated 100 noise+source realizations
for the best-fit number counts model described in
\S~\ref{nc_parametric}.  For each realization, we find the number of
positive and negative ``detections'' just as for the true map.  False
positives are defined as detections occurring $>$10\,arcsec away from
{\em inserted} sources of brightness $>$0.1\,mJy.  The FDR results for
these simulations are given in rows 3 and 6 of Table~\ref{table_fdr}.
The results show that the negatives rate is indeed boosted by the
presence of sources, compared to pure noise maps (rows 2 and 5).
Furthermore, the negative FDR of the actual map (row 4), which
drops to 0 at a S/N of 4.2, is statistically consistent with the
simulated negative FDR means of row 6.  As expected, the simulated
false positives rate is lower than the pure-noise FDR, as
evident from row~3.

As the true positive FDR depends on the number counts, we adopt the
model-independent pure-noise values of row 2 as our nominal FDRs.
These will be conservative overestimates of the FDR regardless of the
true $1.1\,$mm number-counts of the GOODS-N field.

Based on these nominal FDRs, we divide the source candidate list of
table~\ref{table_sources} into two categories of
robustness, with the dividing line at a S/N of 3.75.  On average, we
expect 1--2 source candidates with S/N$\geq$3.75 (above the horizontal
line in Table~\ref{table_sources}) and 1--3 candidates with S/N$<$3.75
(below the line) to be false detections.

\subsection{Flux bias correction}
\label{sources_deboost}

In our map, where the signal from sources does not completely dominate
over noise, the measured flux density can be significantly shifted
from the true $1.1\,$mm flux density of a source due to noise.  The
measured flux densities in column~4 of Table~\ref{table_sources} are
more likely to be overestimates than underestimates of the true flux
densities because of the sharply decreasing surface density of (sub)mm
galaxies with increasing flux density.  As this slope in the number
counts is quite steep \citep[see for example][]{Blain1999A, Barger1999A,
Eales2000, Borys2003, Greve2004, Coppin2006}, this {\em bias} can be a
large effect.  Therefore, we estimate a ``de-boosted'' flux density
for all our 3.5-$\sigma$ source candidates.  This estimate is based on
the Bayesian technique laid out in
\citet{Coppin2005} for calculating the posterior flux density (PFD)
distribution of each source.

The number-counts model that we use to generate the prior is given by
\begin{equation}
{\mathrm{d}N \over \mathrm{d}S} = N^\prime {S^\prime \over S} e^{-S/S^\prime}
\label{eq_prior}
\end{equation}
where d$N$/d$S$ represents the differential number counts of sources
with flux density $S$.  We use $N^\prime=3500\,$mJy$^{-1}$deg$^{-1}$
and $S^\prime=1.5\,$mJy, which is consistent with taking the Schechter
function number-counts fit of
\citet{Coppin2006} and scaling the 850$\,\mu$m fluxes by a factor of 2.2
to approximate the $1.1\,$mm fluxes of the same population.  It is
sufficient to use a prior that is only approximately correct, since
many of the derived results (as we have checked explicitly) are
independent of the exact form of the assumed number counts.  We take
as our Bayesian prior the noise-less pixel flux histogram of a large
patch of sky simulated according to this model.  Since our
point-source kernel has a mean of zero, the prior is non-zero for
negative fluxes and peaks near $0\,$mJy.

The de-boosted flux density given in column~6 of
Table~\ref{table_sources} is the location of the PFD's local maximum
closest to the measured flux density.  This de-boosted flux density is
fairly insensitive to changes in the prior that correspond to other
number-counts models allowed by current constraints.  The upper and
lower error bounds quoted for a de-boosted flux density correspond to
the narrowest PFD interval bracketing the local maximum that
integrates to 68.3\%.

In order to determine the relative robustness of each source
individually, we calculate the integral of the PFD below zero flux.
This quantity, given in column~7 of Table~\ref{table_sources}, is not
a function of just S/N but depends on the flux (signal) and its error
(noise) separately.  Although the values given in column~7 can vary
appreciably among reasonable choices of number-counts priors and the
PFD integration upper bounds (set to zero here), the source robustness
{\it order\/} inferred by the non-positive PFD integral is quite
insensitive to these choices.  Therefore, the values in column~7
provide a useful indicator of the relative reliability of individual
sources.

However, due to the arbitrariness present, the values in
this column cannot be used to directly calculate the FDR of a source
list.  For instance, the sum of column~7 values for our robust source
list is $\sim$0.5, which is an underestimate of the expected FDR (see
\S~\ref{sources_fdr}).  We note that, for our choice of prior, the
requirement of a non-positive PFD $\le$5\% {\em happens to} identify
the same robust source-candidate list as the S/N cut of 3.75.
However, this statement is specific to a particular choice of prior
and PFD integration upper bound.

\subsection{Survey completeness and comparison with SCUBA detections}
\label{sources_completeness}

\begin{figure}
\centering
\includegraphics[width=2.4in,angle=90]{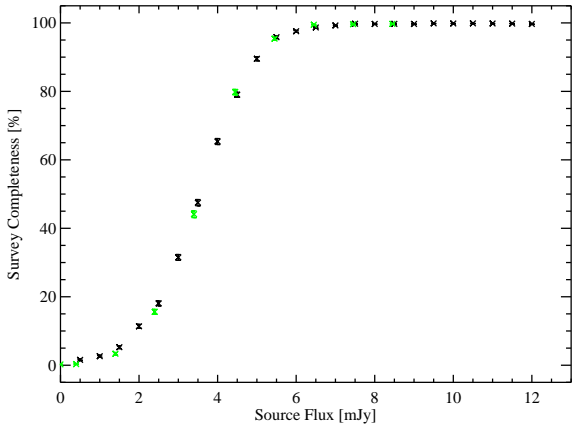}
\caption{Survey completeness for the S/N$\geq$3.75 cut used here to
select robust sources is represented with the dark symbols and error
bars.  The lighter symbols and error bars are estimates of the survey
completeness when the integrated posterior flux distribution below
0\,mJy is required to be $<$5\%.}
\label{fig_completeness}
\end{figure}
We next compute the survey completeness by injecting one source at a
time, in the form of the point-source kernel scaled to represent each
flux, at random positions in the GOODS-N signal map
(Fig.~\ref{fig_source_map}) and tallying the instances when a {\em
new\/} source is recovered with S/N$\geq$3.75 within 10\,arcsec of the
insertion point. We choose this radius because it is small enough for
conducting quick searches in our simulations and because, barring
incompleteness, simulations show that $>$99.5\% of $\ge3.75$-$\sigma$
sources will be found within 10\,arcsec of their true position given
the size of the AzTEC beam and the depth of coverage. This method of
calculating completeness allows for the inclusion of ``confusion
noise'' without altering the map properties appreciably, because only
one artificial source is injected per
simulation~\citep{sescott2006,Scott08}.

We have also assessed completeness by inserting point sources of known
flux, one at a time, into pure noise-realisation maps rather than the
signal map.  With this method, we also require that each recovered
artificial source has a $<5$\% non-positive PFD.  Since this
constraint is essentially equivalent to a limiting S/N threshold of
3.75 (as evident from Table~\ref{table_sources}), it is not surprising
that the survey completeness determined this way (lighter-shade points
of Fig.~\ref{fig_completeness}) is similar to that derived from the
previous method.  The similarity in results also shows that the
effects of confusion noise on survey completeness is small.

Finally, to verify that our source-candidate list has overlap with
previously detected extragalactic (sub)mm sources, we compare our
source list against 850-\um\ SCUBA detections within overlapping
survey regions.  For this purpose, we only consider the regions in the
SCUBA/850-\um\ map with noise r.m.s. $<$ 2.5\,mJy.  Of the 28 AzTEC
sources in the robust list, 11 lie within this region of the SCUBA
map; of these 8 (73\%) have robust detections at 850\,\um\
\citep{Pope2005} within 12\,arcsec of the AzTEC position.  Those 8 are
highlighted with the superscript ``$S$'' in Table~\ref{table_sources}
while the other 3 are marked with the superscript ``$N$.''  On
the other hand, all 38 robust SCUBA sources within the r.m.s. $<$
2.5\,mJy region
\citep{Pope2005,Wall2008} lie within the 70\% coverage region of
AzTEC.  In Chapin et al. (in prep.) we will
discuss the 850\,\um\ properties of AzTEC sources by performing
photometry in the SCUBA map at AzTEC positions, and more fully explore
the overlap of the AzTEC and SCUBA populations in general.

\section{1.1~mm Number counts}
\label{nc}

Using our AzTEC/GOODS-N data, we next quantify the number density of
sources as a function of their intrinsic (de-boosted) 1.1\,mm flux.
These counts cannot be read directly from the recovered distribution
of source flux densities due to: 1) the bias towards higher fluxes in
the data (as described in section~\ref{sources_deboost}), which
includes false detections; and 2) the survey incompleteness at lower
fluxes.  In order to estimate the counts we use two independent
methods: a Monte Carlo technique that implicitly includes the flux bias
and completeness issues; and a Bayesian approach that accounts for
both these effects explicitly.

Fig.~\ref{fig_diffnc_results} shows the results of our number-counts
simulations.  It shows the source flux density histogram simulated for
the best fit model from the parametric method overlaid on the actual
distribution from the true map.  It also shows the differential number
counts vs.\ de-boosted source flux density as returned by both
methods.  The dot-dashed lines in the lower right correspond to the
survey limits of the frequentist and Bayesian approaches, which are
27.8 and 33.8\,deg$^{-2}$\,mJy$^{-1}$, respectively.  The survey limit
is the y-axis value (number counts) that experiences Poisson
deviations to zero sources per mJy-bin 32.7\% of the time, given the
map area considered.  The two limits differ slightly because the
frequentist simulations include the slightly larger area 50\% coverage
region, as opposed to the 70\% coverage region that we use for the
Bayesian method.  The survey limit occurs at around 6\,mJy for both
the best-fit frequentist and Bayesian type simulations.  Thus, we are
not sensitive to the differential number counts {\em with 1\,mJy
resolution\/} beyond that point.

The power of the AzTEC/GOODS-N survey is in constraining number-counts
at lower flux densities, given the depth reached in this relatively
small field.  We have, however, excluded results below the $<$2\,mJy
level from both methods, because of low survey completeness ($<$10\%)
and the possibility of increasing systematic effects.  Therefore, the
noteworthy features of Fig.~\ref{fig_diffnc_results} are the points
from the Bayesian approach, indicated by crosses and error bars, in
the range 2\,mJy to 6\,mJy and the allowed functional forms from the
parametric (frequentist) method within those flux density bounds.
Models allowed by the 68.3\% confidence interval of the parametric
method form the shaded region while the dark curve is the best-fit
model.  Given the error bounds from the two methods, they are in good
agreement.  Both methods are briefly described below.

\subsection{Parametric frequentist approach}
\label{nc_parametric}

An obvious choice of indicator for the underlying source population is
the recovered brightness distribution of source candidates in the
GOODS-N map.  Here, we use a S/N threshold of 3.5 and the 50\%
coverage region of the map.  After identifying S/N$\geq$3.5 source
candidates, we make a histogram of their measured flux densities using
0.25\,mJy bins, for comparison against histograms made from simulating
various number-counts models.  This approach is similar, in spirit, to
the method employed in \citet{Laurent2005} and the parametric version
of number counts derived in \citet{Coppin2006}.  However, we avoid
intermediate analytical constructs, as the procedure outlined below
accounts for all relevant effects.

\begin{figure}
\centering
\includegraphics[width=2.45in, angle=90]{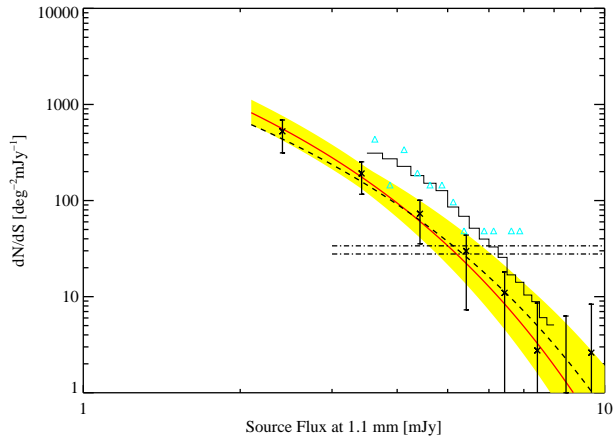}
\caption{The thick solid curve and the enveloping shaded region
correspond to the best fit number counts model and the 68.3\%
confidence interval from the parametric approach of
\S~\ref{nc_parametric}.  The distribution of measured fluxes of
3.5-$\sigma$ sources in the actual map is shown by the triangles in
the 3.5-8\,mJy interval while the corresponding average distribution
of the best fit model is indicated by the thin solid-line histogram.
The difference between the thick solid line and the thin solid
histogram indicates the importance of accounting for flux boosting and
completeness.  The crosses and error bars represent the differential
number counts derived from the Bayesian method, which are in excellent
agreement with the result from the parametric method.  The dashed-line
curve indicates the Bayesian prior.  The upper and lower dot-dashed
lines indicate the survey limits of the Bayesian and parametric
methods, respectively.}
\label{fig_diffnc_results}
\end{figure}

We generate model realisation maps by injecting kernel-shaped point
sources into noise realisation maps.  The input source positions are
{\em uniformly} distributed over the noise realisation map while their
number density and flux distribution reflect the number-counts model
being considered.  For every model we have considered, we make 1200
simulated maps by constructing 12 different source realisations for
each of the 100 noise realisation maps.  Next, we use the same
source-finding algorithm used on the signal map to extract all
S/N$\geq$3.5 peaks in each simulated map.  We then compare the average
histogram of recovered source fluxes from the 1200 model realisations
against the actual distribution of source fluxes.  The data vs.\
models comparison is restricted to the 3.5--8\,mJy measured flux
density range.  This comparison process is illustrated in
Fig.~\ref{fig_diffnc_results}.

The likelihood of the data given a model is determined according to
Poisson statistics as in
\citet{Laurent2005} and \citet{Coppin2006}.
One set of parameterised models that we have explored has the functional form given
by Equation~\ref{eq_prior}.  We chose to re-parametrise these models
so that the normalisation factor depends on only one of the fit
parameters.  The parameters we chose are the same $S^\prime$ as in
Equation~\ref{eq_prior} and $N_{\rm 3mJy}$, the differential counts at
3\,mJy, given by
\begin{equation}
N_{\rm 3mJy} = N^\prime \left({S^\prime \over 3\mathrm{mJy}}\right)
                    e^{-3\mathrm{mJy} / S^\prime}.
\label{eq_N3mJy}
\end{equation}
In terms of these parameters, Equation~\ref{eq_prior} becomes
\begin{equation}
{\mathrm{d}N \over \mathrm{d}S} = N_{\rm 3mJy}
                                  \left({3\mathrm{mJy}} \over S \right)
				  e^{-(S - 3\mathrm{mJy}) / S^\prime}.
\label{eq_ncparam}
\end{equation}

We explored the $S^\prime$--$N_{\rm 3mJy}$ parameter space over the
 rectangular region bracketed by 0.5-2\,mJy and
 60-960\,mJy$^{-1}$\,degree$^{-2}$ using a $(\Delta S^\prime,
\Delta N_{\rm 3mJy})$ cell size of (0.15,60).  The likelihood function, ${\cal
L}$, is a maximum for the model with $S^\prime = 1.25 \pm 0.38\,$mJy
and $N_{\rm 3mJy} = 300 \pm 90$\,mJy$^{-1}$\,degree$^{-2}$.  We did
not assume $\chi^2$-like behaviour of $-\ln({\cal L})$ for calculating
the 68.3\% confidence contours whose projections are the error bars
quoted above.  Instead, as outlined in \citet{press92}, we made many
realisations of the best-fit model and put them through the same
parameter estimation procedure that was applied to the actual data.
In terms of the goodness of fit, we find that 66\% of the simulated
fits yield a higher value of $-\ln({\cal L})$ compared to the actual
value.  Fig.~\ref{fig_diffnc_results} shows this best-fit
number-counts estimate against the de-boosted 1.1\,mm flux density
along with a continuum of curves allowed by the 68.3\% confidence
region.

\subsection{Bayesian method}
\label{nc_bayes}

We also estimate number counts from the individual source PFDs
calculated in \S~\ref{sources_deboost} using a modified version of the
bootstrapping method described in \citet{Coppin2006}.  A complete
discussion of the modifications and tests of the method will be
presented in Austermann et al. (in prep.).  For these calculations, we
use only the sub-list of {\em robust\/} sources in
Table~\ref{table_sources}.  We have repeated this bootstrapping
process 20{,}000 times to measure the mean and uncertainty
distributions of source counts in this field.  The differential and
integrated number counts extracted with this method, using $1\,$mJy
bins, are shown in Fig.~\ref{fig_diffnc_results}.  Our simulations
show that the extracted number counts are quite reliable for a wide
range of source populations and only weakly dependent on the assumed
population used to generate the Bayesian prior (the dashed-line curve
of Fig.~\ref{fig_diffnc_results}) with the exception of the lowest
flux density bins, below 2\,mJy, which suffer from source confusion
and low (and poorly constrained) completeness.  Overall, the results
from the Bayesian method are in excellent agreement with those from
the parametric method between the lower sensitivity bound (2\,mJy) and
the survey limit ($\sim$6\,mJy).

\section{Discussion}
\label{results}

\begin{figure}
\centering
\includegraphics[width=\hsize]{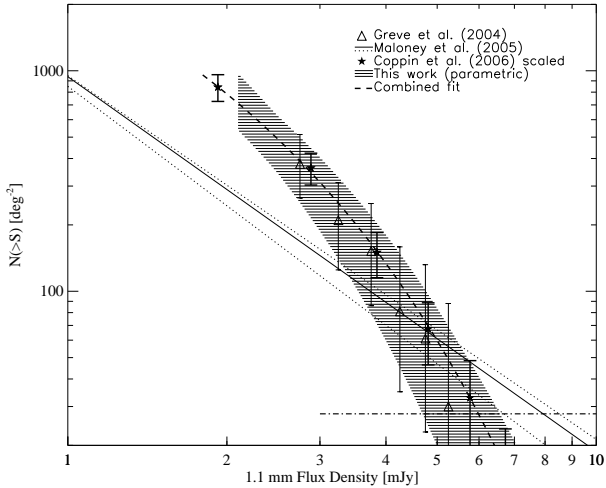}
\caption{The cumulative (integral) number counts from other
1.1--1.2\,mm surveys are shown alongside our results.  The
AzTEC/GOODS-N parametric number-counts results are indicated by the
hatched region that represents the 68.3\% confidence region for
parametric models.  The dot-dashed line indicates the survey limit.
Results from the Bolocam 1.1\,mm Lockman Hole survey are indicated by
a thin solid line and two bounding dotted-lines that represent the
best-fit model and 68.3\% confidence region as found by
\citet{Maloney2005}.  The 1.2-mm MAMBO-IRAM results reported in
\citet{Greve2004} also shown (triangles).  The stars represent the
``reduction D'' results of \citet{Coppin2006} with 850\,\um\ flux
densities scaled by the factor 1/2.08 as explained in the text.  The
dashed curve indicates the best combined fit to the Bayesian results
from both surveys.}
\label{fig_int_counts}
\end{figure}
In Fig.~\ref{fig_int_counts}, we display our cumulative number-counts
results with the 68.3\%-allowed hatched region derived from the
parametric method.  We next compare those results with previous
surveys of (sub)mm galaxies.  Combined results from the 1.2-mm MAMBO
surveys of the Lockman Hole and ELAIS-N2 region~\citep{Greve2004} and
the 1.1-mm Bolocam Lockman Hole survey~\citep{Maloney2005} are shown
in Fig.~\ref{fig_int_counts}.  Our GOODS-N number counts are in good
agreement with MAMBO results.  Our results are in disagreement with
the results of \citet{Maloney2005}, even within a limited flux range
such as 3--6\,mJy where we expect both surveys to be sensitive to the
number counts.

In Fig.~\ref{fig_int_counts}, we also compare our results with the
850-\um\ number counts of \citet{Coppin2006}.  If the 1.1--1.2\,mm
surveys detect the same population of sub-mm sources seen by
SCUBA at 850$\,\mu$m -- an assumption that is not obviously valid
given the possible redshift-dependent selection effects
\citep{Blain2002} -- we would expect a general correspondence between
number counts at these two wavelengths, with a scaling in flux density
that represents the spectral factor for an average source.  Therefore,
we perform a simultaneous fit to the SCUBA/SHADES and AzTEC/GOODS-N
{\em differential} Bayesian number counts in order to determine the
average dust emissivity spectral index, $\alpha_{\rm{dust}}$ (and thus
the flux density scaling factor from 1.1~mm wavelength to 850\,\um\
wavelength), and the parameters, $N_{\rm 3mJy}$ and $S^{\prime}$, of
Equation~\ref{eq_ncparam}.  This fit results in the best-fit parameters and
correlation matrix given in Table~\ref{tab:params}.  We overlay the
\citet{Coppin2006} number counts on Fig.~\ref{fig_int_counts} with
the 850-\um\ fluxes scaled by the scaling factor derived from this
fit, which is 2.08$\pm$0.18.  For visual comparison, the shaded region
of Fig.~\ref{fig_int_counts}, which represents our {\em parametric}
result, is sufficient because it represents well the results from both
methods (see Fig.~\ref{fig_diffnc_results}).
Fig.~\ref{fig_int_counts} shows that the scaled SCUBA-SHADES
points fall well within the bounds allowed by our results.
\begin{table}
\centering
\begin{tabular}{r|c|c|c}
\hline
Survey & $S^{\prime}$ & $N_{\rm 3mJy}$ & $\alpha_{\rm{dust}}$ \\
\hline
\hline
AzTEC/GOODS-N &  $1.25\pm0.38$ &  $300\pm90$  &  \\
AzTEC/GOODS-N \\+ SCUBA/SHADES &   $1.60\pm0.25$   &  $274\pm54$  & $2.84\pm0.32$ \\
\hline
 & $S^{\prime}$ &  $N_{\rm 3mJy}$ & $\alpha_{\rm{dust}}$ \\
$S^{\prime}$          &  1   &  0.05       &  -0.32  \\
$N_{\rm 3mJy}$         & 0.05 &  1      &  -0.8  \\
$\alpha_{\rm{dust}}$  & -0.32    & -0.8     &  1  \\
\end{tabular}
\caption{Best-fit Schechter function parameters and dust emissivity 
         spectral index using the Bayesian results from the
         AzTEC/GOODS-N, SCUBA/SHADES, and combined surveys.  The
         correlation matrix for the combined fit is also listed.
         Caveats on this analysis are given in the text.}
\label{tab:params}
\end{table}

The $\alpha_{\rm{dust}}$ of Table~\ref{tab:params} was computed for
the nominal AzTEC and SCUBA band centres, which are 1.1\,mm and
850\,\um\ respectively.  However, the quoted error on
$\alpha_{\rm{dust}}$ brackets the effects of small shifts in the
effective band centres due to spectral index differences between SMGs
and flux calibrators.  The dust emissivity spectral index may also be
estimated by averaging the 1.1\,mm to 850\,\um\ flux density ratio of
individual sources or by performing the appropriate stacking analysis.
Due to the moderate S/N of sources in our surveys, the effects of flux
bias and survey completeness must be accounted for in such analyses.
Therefore, performing a combined fit to the differential number counts
vs.\ de-boosted flux from the two surveys, where those effects are
already included, is an appropriate method for estimating the spectral
index.  From Fig.~\ref{fig_int_counts}, the hypothesis that
SCUBA and AzTEC detect the same underlying source population appears
plausible.

However, we do not comment on the formal goodness of fit as the
$\chi^2$ obtained for the combined fit is unreasonably small because
the full degree of correlation between data points is underestimated
in the standard computation of the two covariance matrices
\citep{Coppin2006}.  In addition, the best-fit parameters of the
combined fit may have a large scatter from a global mean value (if one
exists), due to sample variance, as the two surveys cover different
fields.  Although SCUBA 850\,\um\ number-counts are available for
GOODS-N \citep{Borys2003}, the survey region (see
Fig.~\ref{fig_goodsn_cover}) and the method used to estimate number
counts in that work are quite different from those used here.
Therefore, we chose to fit to the SCUBA/SHADES number counts
\citep{Coppin2006} instead, since they were determined using methods
similar to ours.

\section{Conclusion}
\label{conclusion}

We have used the AzTEC instrument on the JCMT to image the GOODS-N
field at 1.1~mm.  The map has nearly uniform noise of
0.96--1.16$\,{\rm mJy}\,{\rm beam}^{-1}$ across a field of $245\,{\rm
arcmin}^2$.  A stacking analysis of the map flux at known radio source
locations shows that any systematic pointing error for the map is
smaller than 1\,arcsec in both RA and Dec.  Thus, the dominant
astrometric errors for the 36 source-candidates with S/N$\geq$3.5 are
due to noise in the centroid determination for each source.  Using a
S/N$\geq$3.75 threshold for source robustness, we identify a subset of
28 source candidates among which we only expect 1--2 noise-induced
spurious detections.  Furthermore, of the 11 AzTEC sources that fall
within the considered region of the SCUBA/850-\um, 8 are detected
unambiguously.

This AzTEC map of GOODS-N represents one of the largest, deepest
mm-wavelength surveys taken to date and provides new constraints on
the number counts at the faint end (down to ${\sim}\,2\,{\rm mJy}$) of
the 1.1~mm galaxy population. We
compare two very different techniques to estimate the number density
of sources as a function of their intrinsic flux--a frequentist
technique based on the flux histogram of detected sources in the map
similar in spirit to that of \citet{Laurent2005},
and a Bayesian approach similar to that of
\citet{Coppin2006}.  Reassuringly, the two techniques give similar
estimates for the number counts.  Those results are in good agreement
with the number counts estimates of
\citet{Greve2004} but differ significantly from those of \citet{Maloney2005}.

The 1.1~mm number counts from this field are consistent with a direct
flux scaling of the 850~\micron\ SCUBA/SHADES number counts
\citep{Coppin2006} within the uncertainty of the two measurements,
with a flux density scaling factor of $2.08\pm0.18$.  If we assume
that the two instruments are detecting the same population of sources,
we obtain a grey body emissivity index of $2.84\pm0.32$ for the dust
in the sources.  While there is no evidence based on the number counts
that 1.1~mm surveys select a significantly different population than
850~\micron\ surveys, we caution that the number counts alone cannot
really test this hypothesis.  A more thorough study of whether AzTEC
is selecting a systematically different population than SCUBA can come
only from comparison of the redshifts and multi-wavelength SEDs of the
identified galaxies, which we will describe in Chapin et al. (in
prep.), the second paper in this series.

There is also a survey of GOODS-N with MAMBO at 1.25\,mm performed by
Greve et al. (in prep.).  A comparison between these two millimetre
maps and, possibly, the SCUBA `Super-map' is reserved for a future
paper (Pope et al. in prep.).

This AzTEC/GOODS-N map is one of the large blank-field SMG surveys at
1.1~mm taken at the JCMT. Combined with the AzTEC surveys in the
COSMOS \citep{Scott08} and SHADES (Austermann et al. in prep.)
fields, these GOODS-N data will allow a study of clustering and 
cosmic variance on larger spatial scales than any existing (sub)mm
extragalactic surveys.

\section*{Acknowledgements}
\label{acknowledgement}

The authors are grateful to J. Aguirre, J. Karakla, K. Souccar,
I. Coulson, R. Tilanus, R. Kackley, D. Haig, S. Doyle, and the
observatory staff at the JCMT who made these observations possible.
Support for this work was provided in part by the NSF grant AST
05-40852 and the grant from the Korea Science \& Engineering
Foundation (KOSEF) under a cooperative Astrophysical Research Center
of the Structure and Evolution of the Cosmos (ARCSEC).  DHH and IA
acknowledge partial support by CONACT from research grants 60878 and
50786.  AP acknowledges support provided by NASA through the Spitzer
Space Telescope Fellowship Program, through a contract issued by the
Jet Propulsion Laboratory, California Institute of Technology under a
contract with NASA.  KC acknowledges support from the Science and
Technology Facilities Council.  DS and MH acknowledge support from the
Natural Sciences and Engineering Research Council of Canada.


\end{document}